# Perceived brightness and resolution of holographic augmented reality retinal scan glasses

**Maximilian Rutz[1,2*], Pia Neuberger[2,3], Simon Pick[2], Torsten Straßer[1,4]**

[1]Institute for Ophthalmic Research, University of Tübingen, Elfriede-Aulhorn-Straße 7, 72076 Tübingen, Germany

[2]Bosch Sensortec GmbH, Gerhard-Kindler-Straße 9, 72770 Reutlingen, Germany

[3]Aalen University of Applied Sciences, Beethovenstraße 1, 73430 Aalen, Germany

[4]University Eye Hospital Tübingen, Elfriede-Aulhorn-Straße 7, 72076 Tübingen, Germany

**\* Correspondence:**
Maximilian Rutz
maximilian.rutz@student.uni-tuebingen.de

## Abstract

Augmented reality display performance depends strongly on features of the human visual system. This is especially true for retinal scan glasses, which use laser beam scanning and transparent holographic optical combiners. Human-centered approaches allow us to go beyond conventional optical metrology and evaluate display performance as it is perceived in actual augmented reality use cases. Here, we first present a theoretical formula for the retinal scan luminance and ambient contrast ratio calculated from optical powers, wavelengths, field of view, and human pupil diameter. As a promising insight we found that the pupil diameter dependence is beneficial in assimilating the virtual image luminance to the ambient luminance. Second, we designed and performed a psychophysical experiment to assess perceived resolution in augmented reality settings using a fully functional retinal scan glasses prototype. We present the results of the trials and illustrate how this approach can be useful in the further development of augmented reality smart glasses.

## 1    Introduction

Retinal scan ("RS") glasses are near eye displays that use laser beam scanning for retinal image projection. A collimated light beam is projected via an optical combiner and through the human pupil onto the retina (Suzuki *et al.*, 2018). Recent advances in holographic manufacturing have enabled the use of holographic optical elements ("HOEs") as transparent, diffractive optical combiners (Maimone, Georgiou and Kollin, 2017; Wilm *et al.*, 2021). This allows for lightweight augmented reality displays to be integrated into all-day wearable smart glasses with a small formfactor.

Research and development in the field is progressing quickly (Xiong *et al.*, 2021). First efforts have been made to define new parameters like the ambient contrast ratio ("ACR") that are necessary to characterize the performance of augmented reality displays in real visual environments (Fidler *et al.*, 2021; Park and Lee, 2022). We elaborate on these efforts and show that for the special case of RS displays, interactions with the human eye must be considered.

Most conventional display technologies such as LCD, LED, or waveguides emit light from each pixel into a large solid angle. Their emission profiles roughly obey Lambert's cosine law to keep luminance constant across different viewing angles. We use the term "Lambertian" to describe such displays. From Lambertian displays, only a small portion of the emitted light enters the human eye at any given time. In contrast, all light emitted by RS displays is confined to a narrow scanning beam which passes through the pupil. This makes RS displays not only energy efficient but also leads to important consequences for fundamental display parameters like the brightness and resolution.

One challenge is that some conventional measurement devices such as luminance cameras cannot be used directly with RS glasses due to the interaction of the projection system with the human pupil. Instead, the luminance of an RS display must be computed in a certain way to get accurate results. We introduce a theoretically derived formula to calculate the retinal scan luminance and ambient contrast ratio. This formula incorporates various parameters including optical powers, wavelengths, field of view, and human pupil diameter. Building on this formula, we designed and conducted a psychophysical study to measure the perceived resolution of RS glasses in augmented reality settings.

## 2     Materials and Methods

### 2.1    Retinal Scan Glasses

The BML500P RS glasses prototype available for this study (Fig. 1) was developed by Bosch Sensortec and presented at the Consumer Electronics Show 2020 (Evan Ackermann, 2020).

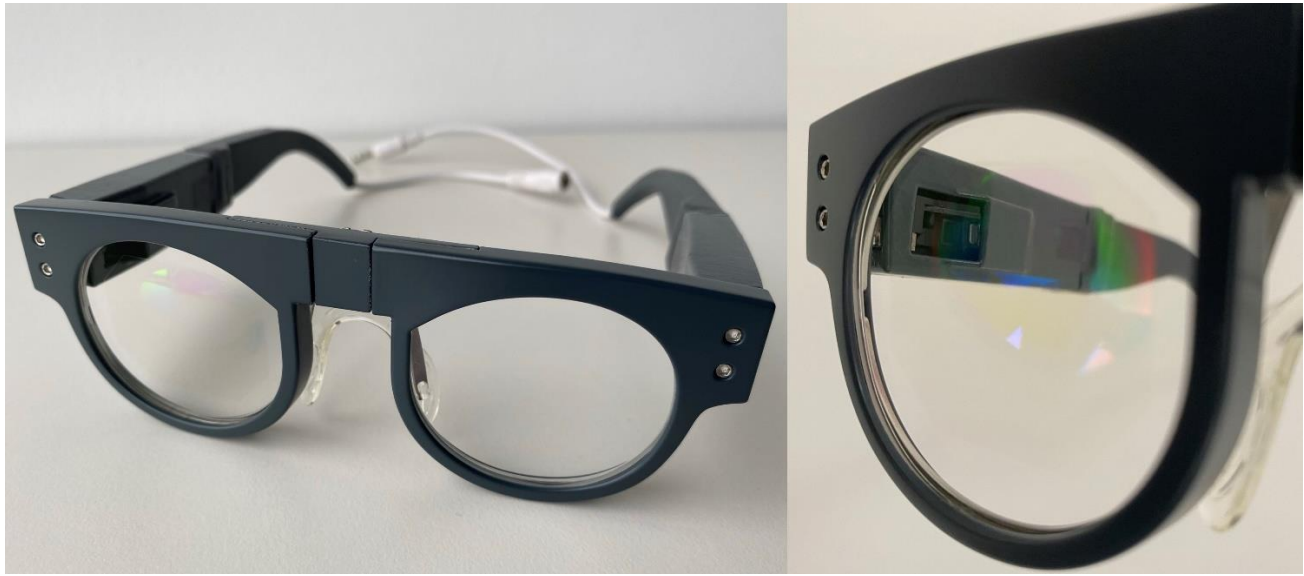

Fig. 1. (Left) BML500P RS glasses prototype with light engine in the right temple. (Right) Transparent holographic optical element embedded in the right spectacle lens.

A light engine in the right temple of the glasses projects images via a transparent holographic optical element through the pupil onto the retina. Two oscillating micromirrors steer light from three low power RGB laser diodes in a raster scanning pattern to form an image at a refresh rate of 60 Hz.

### 2.1.1 Brightness in Retinal Raster Scanning

The human eye detects retinal illuminance as brightness. Retinal illuminance $E_R$ is defined as luminous flux $\Phi_V$ per retinal area $A_R$.

$$E_R = \frac{\Phi_V}{A_R} \tag{1}$$

Display brightness is usually denoted in terms of luminance L, which is defined as the luminous flux $\Phi_V$ emitted into a solid angle $\Omega$ by a light source with area $A_S$.

$$L = \frac{\Phi_V}{\Omega\, A_S} \tag{2}$$

Luminance is a measure for the light source, while the perceived brightness depends on the illuminance on the retina. When comparing Lambertian displays with each other, luminance can be used as a proxy for perceived brightness because luminance changes lead to proportional retinal illuminance changes.

This is however not valid for a comparison between Lambertian and RS displays. The size of the human pupil influences the luminous flux $\Phi_V$ from a Lambertian display but not from an RS display as shown in Fig. 2.

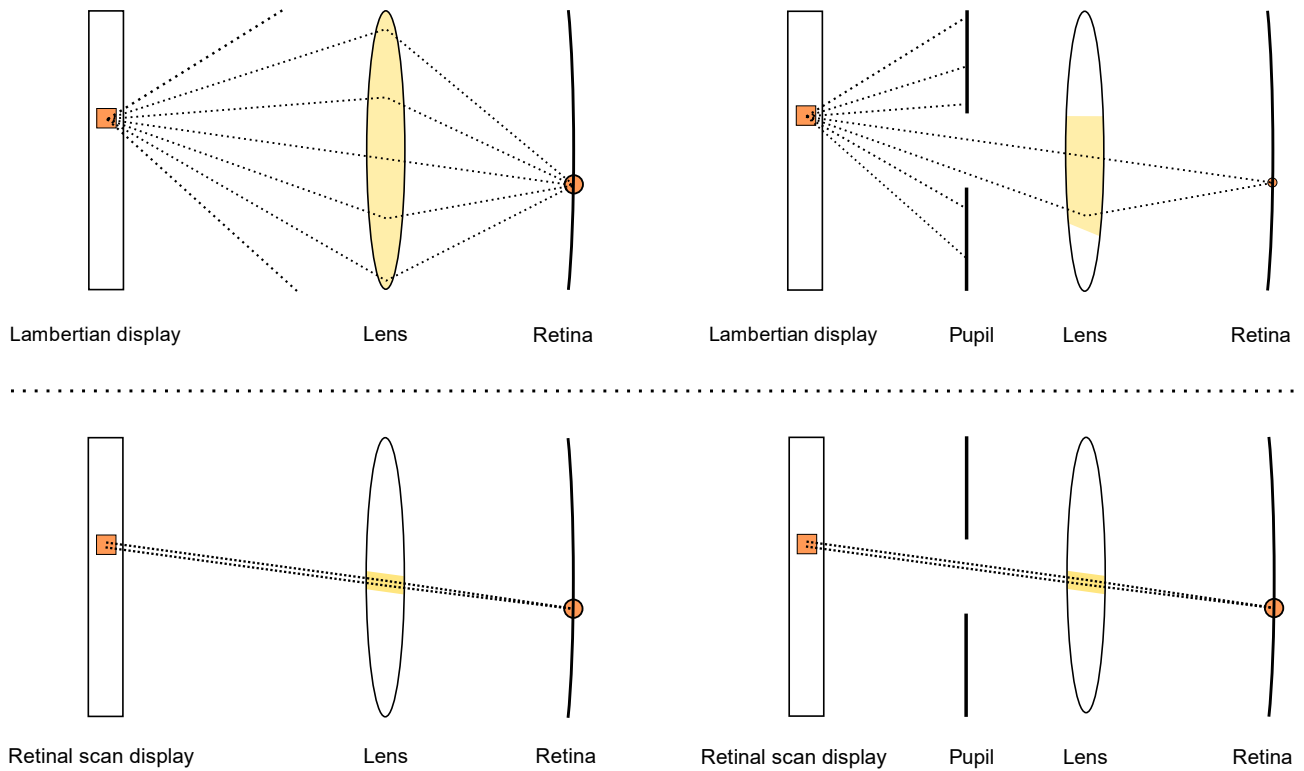

Fig. 2. Beam paths of a single pixel for a Lambertian display (top) and an RS display (bottom). Without an aperture (left) all light rays hitting the lens contribute to the luminous flux on the retina. When the human pupil acts as an aperture (right) the total luminous flux is reduced for a Lambertian display but not for an RS display as long as the narrow RS beam (diameter < 0.5 mm) fully passes through the pupil.

For this reason, measuring the luminance of an RS image with a luminance camera leads to inconsistent results: The larger the entrance pupil of the luminance camera (to which it is calibrated using a Lambertian light source), the smaller the measured RS luminance.

In the first part of the results section, we show that it is nevertheless possible to calculate an equivalent luminance for RS displays. This luminance depends on the RS luminous flux, RS field of view, human pupil size and is directly comparable with the luminance of Lambertian displays.

### 2.1.2 Focus and Resolution in Retinal Raster Scanning

Fig. 3 shows the impact that ocular imperfections have on Lambertian and RS images. The circle of confusion for individual pixels is smaller for RS images due to the strong collimation and long depth of field of the scanning light beam.

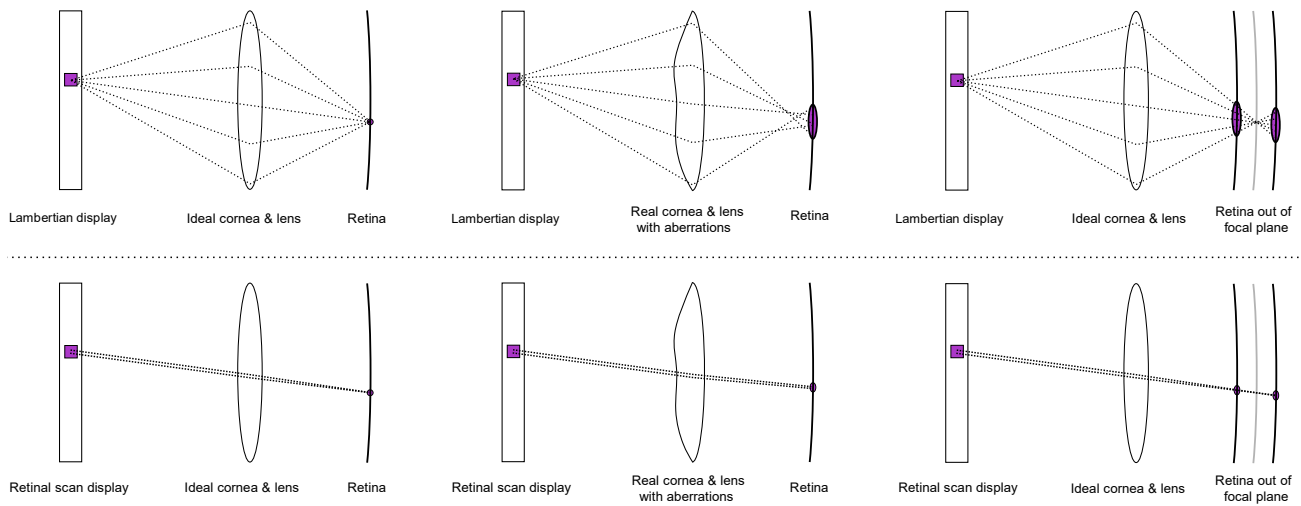

Fig. 3. Beam paths of a single pixel entering the eye for a Lambertian display (top) and an RS display (bottom). An ideal cornea and lens (left) focus light from a single pixel onto a narrow point on the retina. Aberrations of the cornea and lens (middle) or an out-of-focus retinal plane in myopia or hyperopia (right) have a much smaller impact on the focus and perceived resolution of an RS image.

While this effect can be illustrated clearly, it is difficult to assess the differential impact on display performance with methods from conventional display metrology since the human eye needs to be brought "into the loop".

## 2.2 Perceived Resolution Study Design

To measure the perceived resolution of the BML500P RS glasses prototype, we conducted a psychophysical discrimination test with 20 participants (17 males, 3 females; age 18 - 60 years, mean ± sd: 36.0 ± 7.7 years). Inclusion criteria were right-eye dominancy, a right monocular visual acuity of at least 0.8 decimal (screening results: logMAR mean ± sd: -0.073 ± 0.060), a right monocular contrast vision threshold below 15 % Weber contrast (mean ± sd: 4.2 ± 0.8 %) and no known eye diseases. The experiment was approved by the ethics committee of the Medical Faculty of the University of Tübingen (410/2022B02) and conducted according to the provisions of the Declaration of Helsinki. Informed consent was obtained from all participants.

The study setup is shown in Fig. 4. Participants were equipped with the RS glasses on a chinrest 2.3 m in front of a 60-inch LED monitor (Sharp PN-R603). The light-isolated examination room was illuminated at an average of 2500 lux, equal to outdoor illumination on a cloudy day.

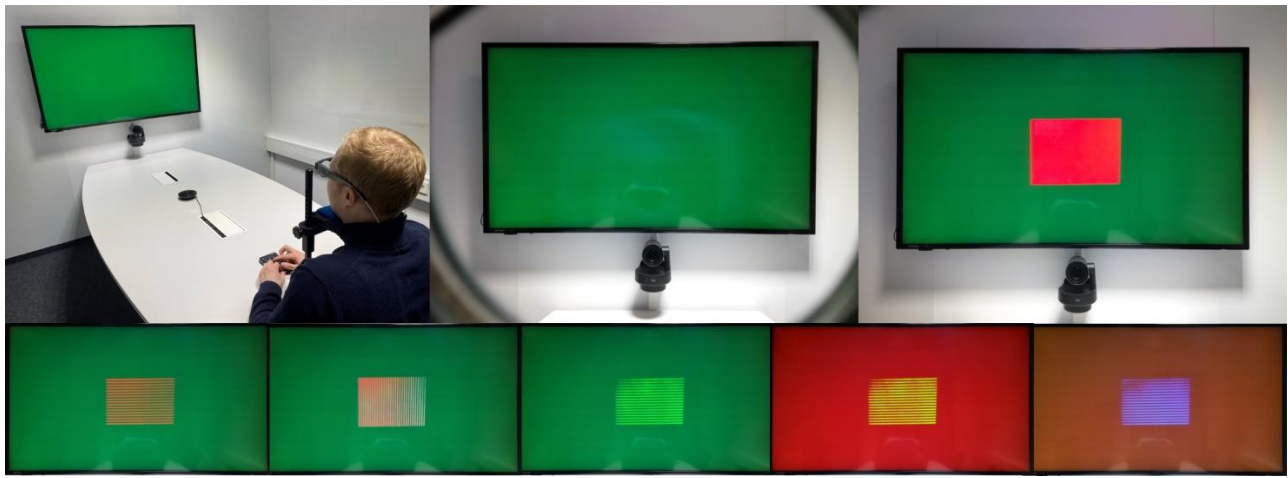

Fig. 4. (Top left) Chinrest in front of LED monitor displaying an even green background. (Top middle) LED monitor seen through right spectacle lens with embedded HOE. (Top right) Red rectangle stimulus from glasses on green background seen through right spectacle lens. (Bottom 1st) Horizontal red lines, green background. (Bottom 2nd) Vertical red lines, green background. (Bottom 3rd) Horizontal green lines, green background. (Bottom 4th) Green horizontal lines, red background. (Bottom 5th) Blue horizontal lines, orange background.

Participants were shown a right-eye monocular stimulus in the glasses with a size of 11.15 ± 0.22 deg x 14.99 ± 0.17 deg [mean ± sd from 10 measurements]. The transparent RS glasses superimposed this virtual image onto the real image from the LED monitor. Psychopy (Peirce *et al.*, 2019) was used to control the monitor and RS glasses stimuli simultaneously. To minimize distractions, participants wore a black eye patch over the left eye.

The glasses stimulus was picked randomly from a selection of monochromatic line patterns with different spatial frequencies from 1.8 cyc/deg to 16.0 cyc/deg and a full monochromatic rectangle. The line pattern consisted of lines with equal digital pixel width for both projection and non-illuminated dark background lines. The spatial frequency range of the lines was determined by the 480 x 320 pixels that could be set within the software configuration of the prototype. The pattern with the highest spatial frequency had a width of one digital pixel.

For high digital spatial frequencies, the resulting optical lines started to blur into a full rectangle. The task of the participants was to report whether they saw lines or a rectangle using a wireless keyboard. The spatial frequency of the lines was adjusted based on the participants' responses using a 3-down-1-up staircase algorithm to determine the highest resolvable spatial frequency. The algorithm terminated after three reversals. There was no time limit for responses. To ensure that a staircase level could not be passed by only identifying rectangles, a line pattern was always presented after two correct responses. A 0.5 second pause between stimuli was added to avoid afterimages and to reduce cues from stimulus transitions.

We performed 15 trials each for horizontal and vertical lines for different stimulus colors (laser wavelengths red [640 nm], green [521 nm], blue [452 nm]), stimulus brightness (maximum radiant flux red [272 nW]) green [736 nW], blue [1263 nW], equiluminant radiant flux green [66 nW], blue [1145 nW]) and LED monitor backgrounds [black, red, green, blue, orange] to measure the influence of different stimulus and background conditions on the perceived resolution.

The LED monitor was set to a luminance of 27 cd/m² for each color and 1 cd/m² for the black background (measured with JETI Spectraval 1501 spectroradiometer). The right spectacle lens with the holographic optical combiner had an optical transmittance of $\tau_C = 0.90$.

Generally, the perceived resolution threshold not only depends on the spatial frequency but also on the contrast. If the contrast between stimuli and background is too small, the perceived resolution threshold decreases across spatial frequencies (Chung and Tjan, 2009; Pelli and Bex, 2013). Previous research showed that this influence is persistent up to contrast levels of about 0.5 (Dressier and Rassotv, 1981). Since we wanted to avoid the influence of contrast on the perceived resolution in this study, we needed to ensure that the contrast between the retinal scan stimuli and the Lambertian background stayed above 0.5. As shown in section 2.1.1, however, this contrast can only be calculated correctly if the brightness for both display types is denoted in the same physical unit. We therefore first derive a formula for this calculation in the first part of the results section before presenting the results of the study in the second part of the section.

## 3 Results

### 3.1 Retinal Scan Luminance and Ambient Contrast Ratio

To express the brightness of an RS display in terms of luminance, we determine the luminance of a Lambertian display that would cause the same retinal illuminance $E_R$ as the RS glasses.

$$E_{R_{Lam}} = E_{R_{Scan}} \tag{3}$$

To find the RS retinal illuminance $E_{R_{Scan}}$ we first calculate the luminous flux $\Phi_{V_{Scan}}$ from the radiant flux $\Phi_{E_{Scan}}$, the luminous efficacy constant $K_m = 683 \frac{lm}{W}$ and the luminous efficiency function $V(\lambda)$. In practice, the integral is approximated by a finite sum of measurements.

$$\Phi_{V_{Scan}} = K_m \int V(\lambda) \, \Phi_{E_{Scan}}(\lambda) \, d\lambda \approx K_m \sum_\lambda V(\lambda) \, \Phi_{E_{Scan}}(\lambda) \tag{4}$$

Next, we calculate the illuminated retinal area $A_{R_{Scan}}$ geometrically from the RS field of view angles α and β and the eye's focal distance $f_e$.

$$A_{R_{Scan}} = f_e \, 2 \tan\left(\frac{\alpha}{2}\right) f_e \, 2 \tan\left(\frac{\beta}{2}\right) = 4 \, f_e^{\,2} \tan\left(\frac{\alpha}{2}\right) \tan\left(\frac{\beta}{2}\right) \tag{5}$$

Plugging Equations (4) and (5) into Equation (1) gives us the retinal illuminance for an RS display.

$$E_{R_{Scan}} = \frac{\Phi_{V_{Scan}}}{A_{R_{Scan}}} = \frac{K_m \sum_\lambda V(\lambda) \, \Phi_{E_{Scan}}(\lambda)}{4 \, f_e^{\,2} \tan\left(\frac{\alpha}{2}\right) \tan\left(\frac{\beta}{2}\right)} \tag{6}$$

Next, we express the Lambertian display retinal illuminance $E_{R_{Lam}}$ in terms of the Lambertian display luminance L (Bass, Enoch and Lakshminarayanan, 2010, p. 183/184). For an observer with pupil area $A_P$ at distance R to the display, the solid angle $\Omega$ will be given by the ratio of $A_P$ and $R^2$.

$$\Omega = \frac{A_P}{R^2} \tag{7}$$

Plugging Equation (7) into Equation (2) and rearranging terms yields the luminous flux $\Phi_{V_{Lam}}$ entering the pupil as a function of the Lambertian display luminance $L_{Lam}$, pupil area $A_P$, display area $A_S$ and display distance R.

$$L_{Lam} = \frac{\Phi_{V_{Lam}}}{\frac{A_P}{R^2} A_S} \Leftrightarrow \Phi_{V_{Lam}} = \frac{L_{Lam} \, A_P \, A_S}{R^2} \tag{8}$$

To find the retinal image area $A_{R_{Lam}}$ we use the lens formula from geometrical optics. A display with area $A_S$ at a distance R to the eye creates an image with area $A_R$ at focal distance $f_e$ on the retina. The magnification between $A_S$ and $A_R$ is given by the square of the ratio between image distance $f_e$ and object distance R.

$$A_{R_{Lam}} = A_S \left(\frac{f_e}{R}\right)^2 \tag{9}$$

Plugging Equations (8) and (9) into Equation (1) gives us the retinal illuminance for a Lambertian display:

$$E_{R_{Lam}} = \frac{\Phi_{V_{Lam}}}{A_{R_{Lam}}} = \frac{L_{Lam} \, A_P \, A_S}{R^2 \, A_S \left(\frac{f_e}{R}\right)^2} = \frac{L_{Lam} \, A_P}{f_e^{\,2}} \tag{10}$$

Finally, we can set both retinal illuminances $E_{R_{Lam}}$ and $E_{R_{Scan}}$ equal.

$$\frac{L_{Lam} A_P}{f_e^2} = \frac{K_m \sum_\lambda V(\lambda) \Phi_{E_{Scan}}(\lambda)}{4 f_e^2 \tan\left(\frac{\alpha}{2}\right) \tan\left(\frac{\beta}{2}\right)} \tag{11}$$

After rearranging terms and renaming $L_{Lam}$ to $L_{RS}$ we arrive at Equation (12) for the luminance $L_{RS}$ of an RS display.

$$L_{RS} = \frac{K_m \sum_\lambda V(\lambda) \Phi_{E_{Scan}}(\lambda)}{4 A_P \tan\left(\frac{\alpha}{2}\right) \tan\left(\frac{\beta}{2}\right)} \tag{12}$$

To calculate the pupil area $A_P$ we measure the pupil diameter $d_P$ in mm.

$$A_P = \frac{1}{4} \pi \left(0.001 \, d_{P_{mm}}\right)^2 \tag{13}$$

For an RS display with narrow RGB laser spectral linewidths, we approximate $\sum_\lambda V(\lambda) \Phi_{E_{Scan}}(\lambda)$ in Equation (12) with radiant powers $\Phi_i$ and average wavelengths $\bar{\lambda}_i$.

$$L_{RS} = \frac{K_m \left(V(\bar{\lambda}_{red}) \Phi_{red} + V(\bar{\lambda}_{green}) \Phi_{green} + V(\bar{\lambda}_{blue}) \Phi_{blue}\right)}{\pi \left(0.001 \, d_{P_{mm}}\right)^2 \tan\left(\frac{\alpha}{2}\right) \tan\left(\frac{\beta}{2}\right)} \tag{14}$$

$K_m$ and $V(\lambda)$ are constant and tabularized (CIE JTC 2, 2019). $\Phi_i$, $\alpha$ and $\beta$ must be measured for the RS display. The only remaining variable is the pupil diameter $d_{P_{mm}}$. Changes in pupil diameter can be caused by different factors (Kelbsch *et al.*, 2019). In most circumstances however, the main contributor is the ambient luminance $L_A$. We use this fact to express the RS luminance as a function of ambient luminance instead of pupil size.

Several mathematical models have been developed for this relationship (Watson and Yellott, 2012). We picked the model from Stanley & Davies (Stanley and Davies, 1995) for binocular background adaptation with an adaptive field size a of (60 deg)$^2$. The model allows us to express pupil diameter $d_P$ in mm as a function of ambient luminance $L_A$ in cd/m$^2$ for an average observer.

$$d_{P_{mm}} = 7.75 - 5.75 \left(\frac{\left(\frac{L_A \, 3600}{846}\right)^{0.41}}{\left(\frac{L_A \, 3600}{846}\right)^{0.41} + 2}\right) \tag{15}$$

Plugging Equation (15) into (14) allows us to calculate the RS luminance $L_{RS}$ as a function of the ambient luminance $L_A$.

$$L_{RS} = \frac{K_m\left(V(\bar{\lambda}_{red})\,\Phi_{red} + V(\bar{\lambda}_{green})\,\Phi_{green} + V(\bar{\lambda}_{blue})\,\Phi_{blue}\right)}{\pi\left(0.001\left(7.75 - 5.75\left(\frac{\left(\frac{L_A\,3600}{846}\right)^{0.41}}{\left(\frac{L_A\,3600}{846}\right)^{0.41} + 2}\right)\right)\right)^2 \tan\left(\frac{\alpha}{2}\right)\tan\left(\frac{\beta}{2}\right)} \quad (16)$$

Expressing the RS luminance in terms of the ambient luminance not only eliminates the need for direct pupil size measurements but also simplifies another useful quantity, the "Retinal Scan ambient contrast ratio" $ACR_{RS}$ which is defined by the RS luminance $L_{RS}$, the combiner transparency $\tau_C$ and the ambient luminance $L_A$.

$$ACR_{RS} = \frac{L_{RS}}{\tau_C\,L_A} \quad (17)$$

In non-transparent displays the contrast ratio is determined from the brightest and darkest pixel's luminance. RS displays do not project any light into fully dark pixels. The luminance of the darkest RS display pixel is therefore given by the luminance of the darkest background element within the field of view. This could be measured for a fixed head orientation and homogeneous background in a controlled laboratory setting. For most natural visual scenes however, such a pixel-precise background luminance would be difficult to determine. For those cases we use the average ambient luminance $L_A$ as an estimate.

Since the ambient light passes through the spectacle lens, its luminance is reduced by the transmittance $\tau_C$ of the HOE combiner. Optical power measurements of the RS display are taken at the eye position, so $\tau_C$ does not influence the numerator in Equation (17). The transmittance could however influence the pupil diameter in Equation (15). This will depend on the size of the combiner in comparison to the adaptive field size. Since the HOE region in our glasses prototype was small, and, more importantly, monocular, we neglected this influence. If this does not hold true for other prototypes, $\tau_C$ can be included as an additional coefficient to $L_A$ in Equation (15).

Fig. 5 shows the RS luminance $L_{RS}$ and RS ambient contrast ratio $ACR_{RS}$ as a function of the ambient luminance $L_A$ for a fixed set of RS glasses parameters.

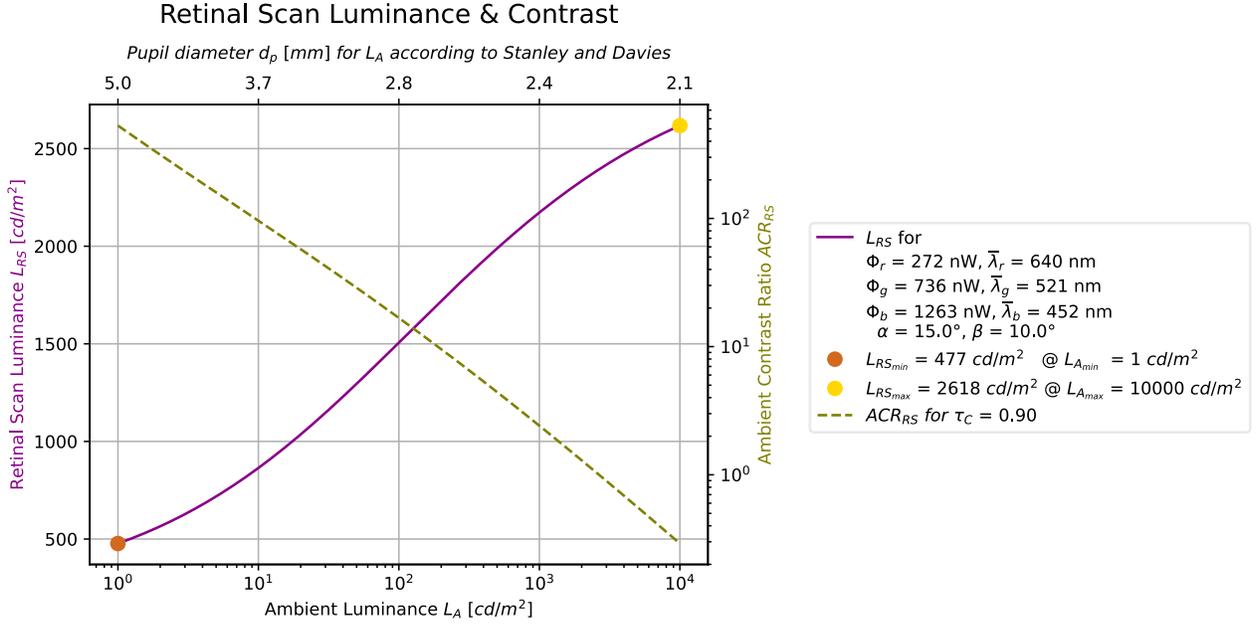

Fig. 5. RS luminance $L_{RS}$ and RS ambient contrast ratio $ACR_{RS}$ for ambient luminance $L_A$ from 1 to 10,000 cd/m² and corresponding pupil diameters $d_p$. Radiant fluxes $\Phi_i$, average wavelengths $\bar{\lambda}_i$, field of view angles α, β and combiner transparency $\tau_C$ are constant.

### 3.2 Perceived Resolution Study Results

The study was conducted in a controlled environment which allowed us to determine the pupil size directly. We measured the pupil diameter of all participants right before the experiment while they faced the monitor on the chinrest and found a mean ± sd pupil diameter of 3.3 ± 0.4 mm within a range from 2.5 mm to 4.0 mm. With the optical measurement values from the glasses, we determined $L_{RS}$ according to Equation (14). We then used the luminance of the LED monitor backgrounds for $L_A$ and the transmittance of the combiner $\tau_C$ to calculate $ACR_{RS}$ according to Equation (17).

**Table 1** provides the trial parameters and results from the perceived resolution study. The left side of the table shows the trial parameters including the ambient contrast ratio $ACR_{RS}$ for the retinal scan stimuli and LED monitor background. Since $ACR_{RS}$ stayed well above 0.5 for all trials we can now assume that the perceived resolution threshold is only a function of the spatial frequency and not the contrast. The right side of the table shows the number of participants that successfully passed each staircase level. "Passing" a staircase level means that the participant provided three correct answers in a row at that level during the staircase algorithm. The level number corresponds to the digital pixel width of the line patterns. All 20 participants were able to complete horizontal level 4H (3.59 cyc/deg) and vertical level 5V (3.20 cyc/deg) in all trials. No participant was able to complete horizontal level 1H (14.35 cyc/deg) or vertical level 1V (16.01 cyc/deg) in any trial.

**Table 1.** Trial parameters and results from staircase algorithm in perceived resolution study

| Trial | Trial Parameters | | | | Number of participants that passed staircase resolution level Horizontal [H] and Vertical [V] | | | | | | | |
|---|---|---|---|---|---|---|---|---|---|---|---|---|
| | Glasses | Monitor | $L_{RS}$ [cd/m$^2$] | $ACR_{RS}$ | 4H *3.59 c/d* | 3H *4.78 c/d* | 2H *7.17 c/d* | 1H *14.35 c/d* | 4V *4.00 c/d* | 3V *5.34 c/d* | 2V *8.01 c/d* | 1V *16.01 c/d* |
| 1 | red | black | 83 | 92 | 20 | 17 | 0 | 0 | 20 | 17 | 0 | 0 |
| 2 | red | green | 83 | 3 | 20 | 19 | 0 | 0 | 20 | 18 | 0 | 0 |
| 3 | red | red | 83 | 3 | 20 | 19 | 0 | 0 | 19 | 19 | 0 | 0 |
| 4 | green | black | 929 | 1032 | 20 | 20 | 4 | 0 | 20 | 20 | 0 | 0 |
| 5 | green | black | 83 | 92 | 20 | 20 | 5 | 0 | 20 | 20 | 0 | 0 |
| 6 | green | red | 929 | 38 | 20 | 20 | 5 | 0 | 20 | 20 | 0 | 0 |
| 7 | green | red | 83 | 3 | 20 | 20 | 6 | 0 | 20 | 20 | 0 | 0 |
| 8 | green | green | 929 | 38 | 20 | 20 | 6 | 0 | 20 | 20 | 0 | 0 |
| 9 | green | green | 83 | 3 | 20 | 20 | 5 | 0 | 20 | 20 | 0 | 0 |
| 10 | blue | black | 91 | 101 | 20 | 20 | 4 | 0 | 20 | 20 | 0 | 0 |
| 11 | blue | black | 83 | 92 | 20 | 20 | 5 | 0 | 20 | 20 | 2 | 0 |
| 12 | blue | orange | 91 | 4 | 20 | 20 | 3 | 0 | 20 | 20 | 1 | 0 |
| 13 | blue | orange | 83 | 3 | 20 | 20 | 4 | 0 | 20 | 20 | 1 | 0 |
| 14 | blue | blue | 91 | 4 | 20 | 20 | 5 | 0 | 20 | 20 | 1 | 0 |
| 15 | blue | blue | 83 | 3 | 20 | 20 | 5 | 0 | 20 | 20 | 1 | 0 |

We additionally fitted sigmoidal psychometric functions to the individual trial data. Fig. 6 shows two separate fits for horizontal and vertical lines. We determined the perceived resolution threshold from the inflection point of the function. This gives us a perceived resolution of 7.1 cyc/deg for horizontal and 7.4 cyc/deg for vertical lines.

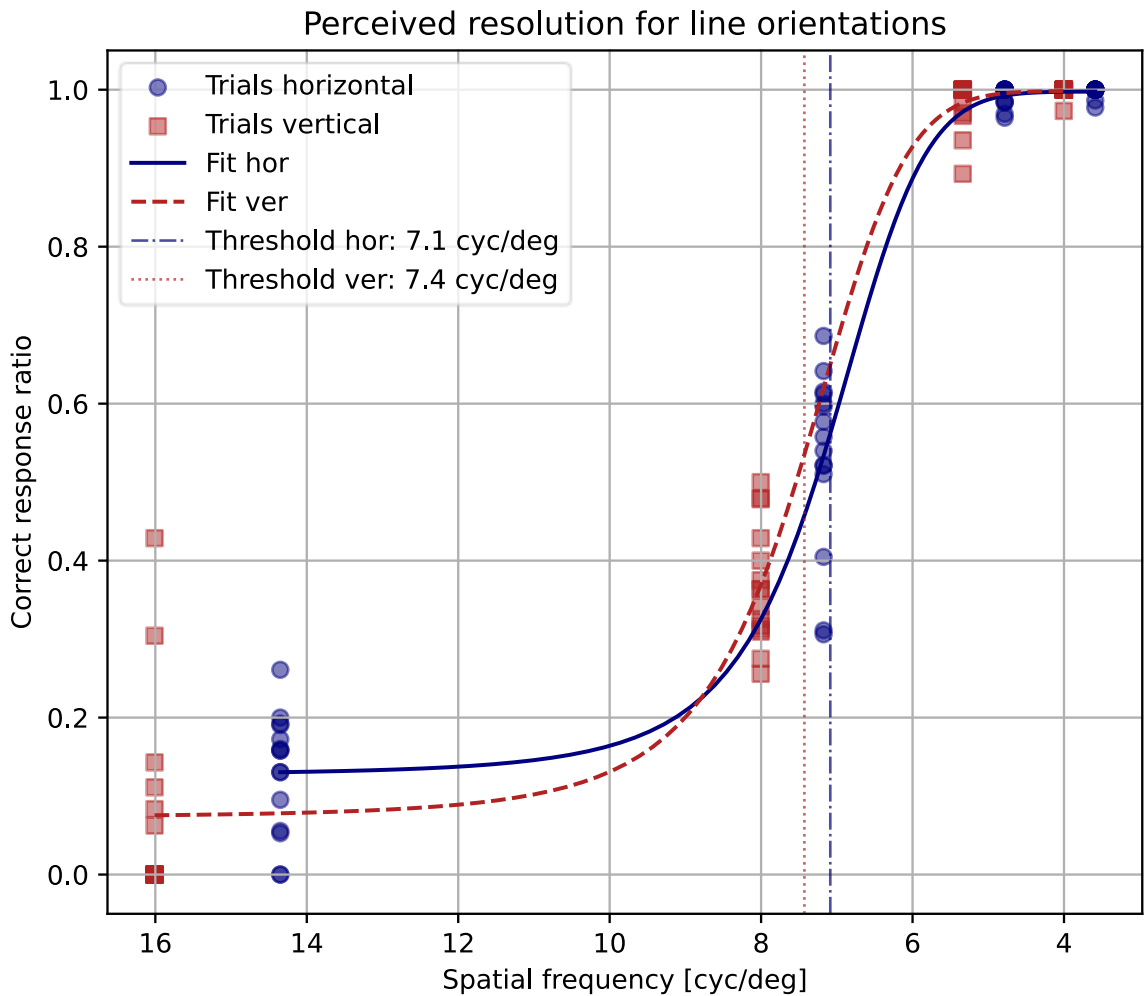

**Fig. 6. Sigmoidal psychometric curves for perceived resolution of horizontal and vertical lines**

We fitted separate psychometric functions for trials with different stimulus and background conditions but found similar thresholds for all of them.

The biggest differences can be seen for trials with different stimulus colors. Participants reached a threshold of 6.7 cyc/deg in trials with red stimuli, a threshold of 7.4 cyc/deg in trials with green stimuli and a threshold of 7.3 cyc/deg in trials with blue stimuli (Fig. 7). These results agree with observations from **Table 1**, since participants performed slightly worse in the staircase algorithm for red stimuli as well.

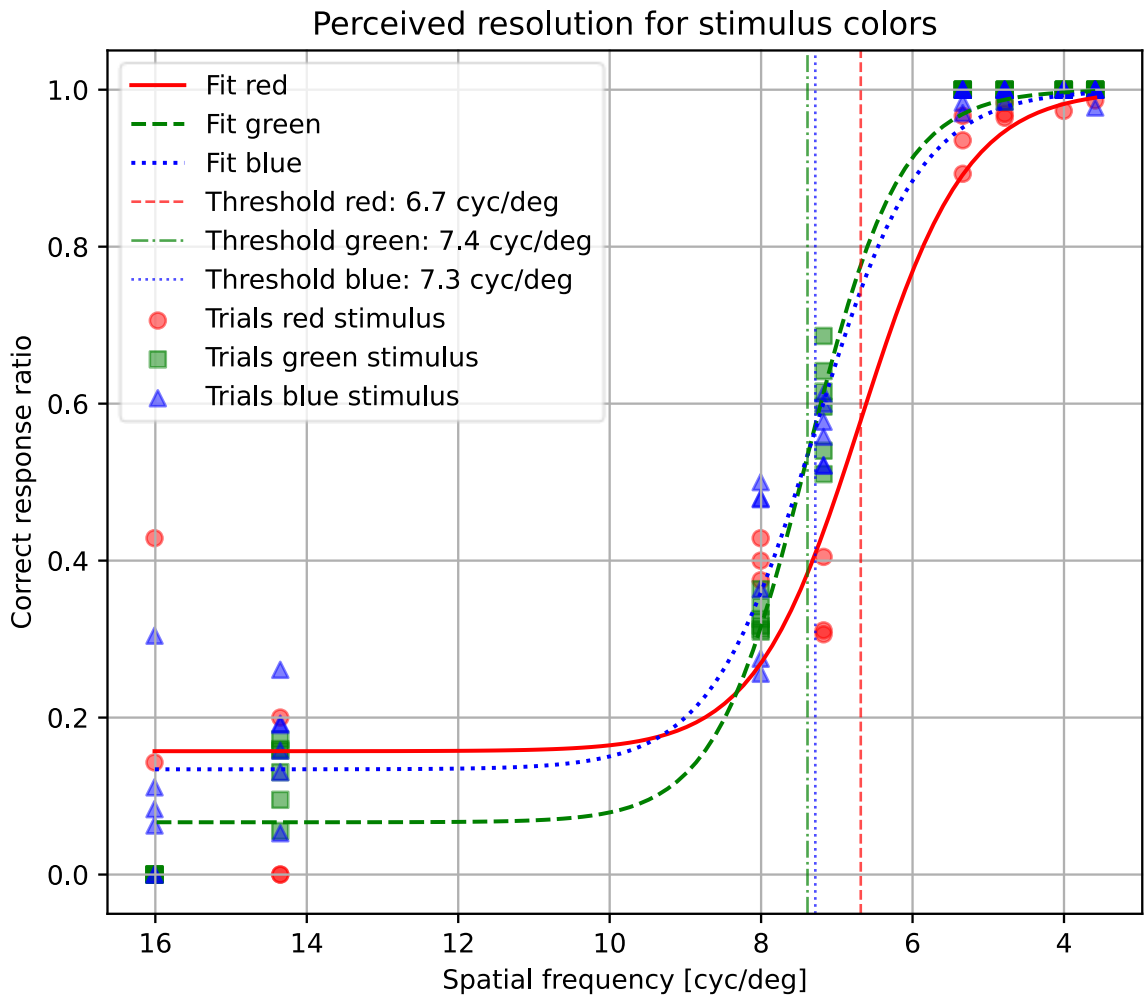

**Fig. 7. Sigmoidal psychometric curves for perceived resolution with different stimulus colors**

Participants achieved a threshold of 7.1 cyc/deg in trials with maximum stimulus luminance and a threshold of 7.4 cyc/deg in trials with equal stimulus luminance (Fig. 8).

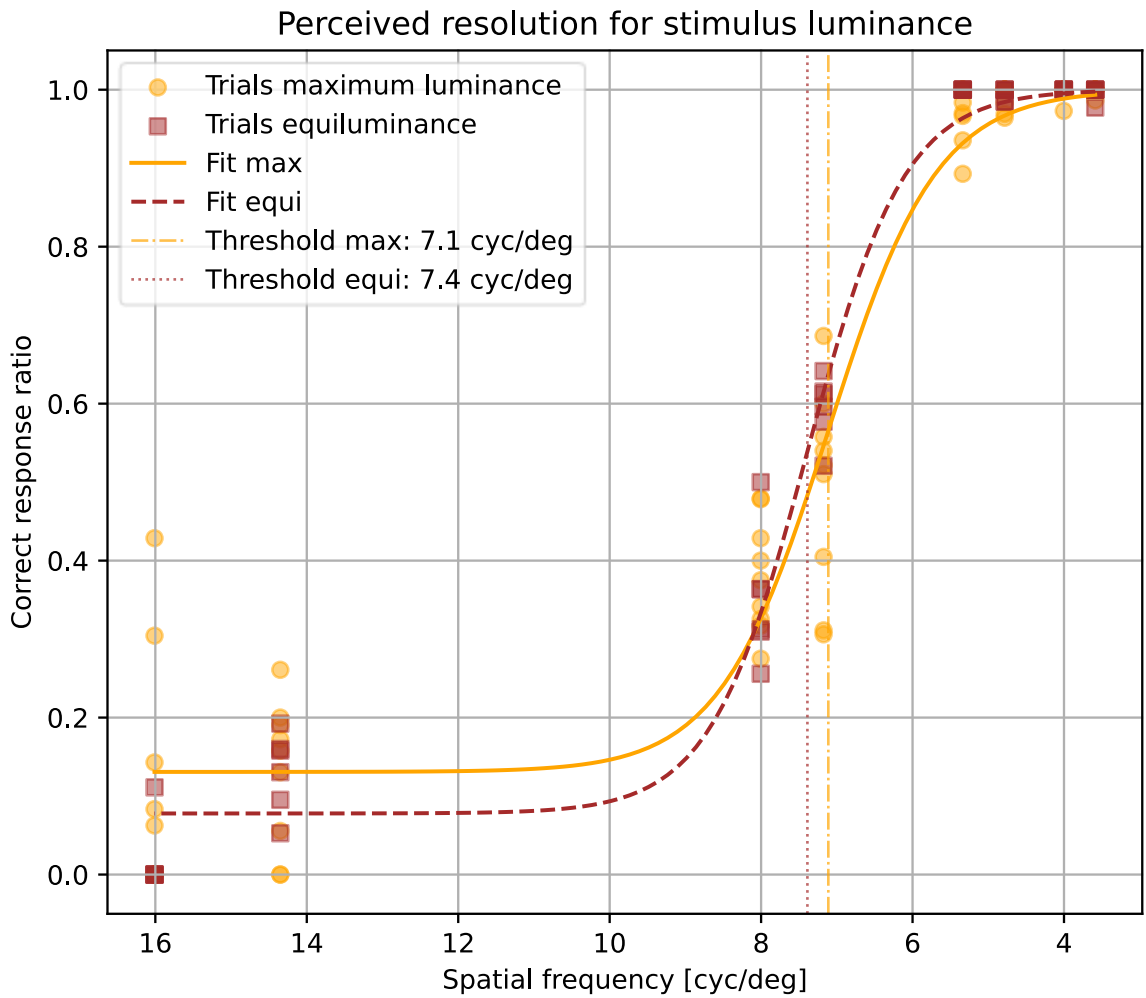

**Fig. 8. Sigmoidal psychometric curves for perceived resolution with different stimulus brightness**

They reached a threshold of 7.2 cyc/deg in trials with a black background, a threshold of 7.2 cyc/deg in trials with complementary color backgrounds (red/green, green/red, blue/orange) and a threshold of 7.4 cyc/deg in trials with the same color backgrounds (Fig. 9).

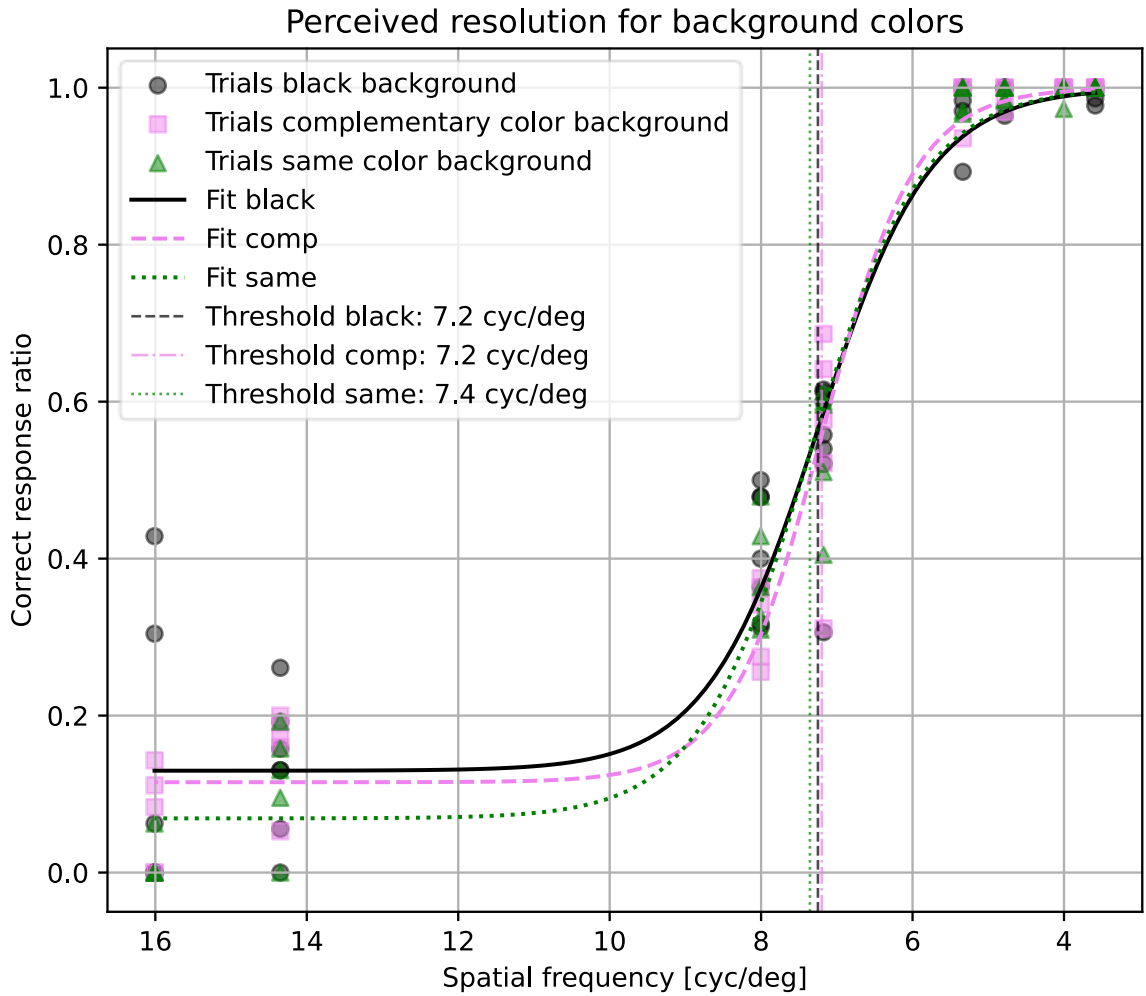

**Fig. 9. Sigmoidal psychometric curves for perceived resolution with different stimulus and background color combinations**

## 4 Discussion

RS glasses use an optical architecture that is different from conventional display technologies. We explained how the human pupil influences the perceived brightness of an RS display and presented a physical derivation for the RS luminance $L_{RS}$. Since the pupil adapts to ambient luminance, RS glasses appear darker in dark environments and brighter in bright environments. For an observer with standard pupil dynamics, the RS display luminance multiplies fivefold from mesopic to bright photopic environments.

For conventional displays, Troland is commonly used to describe retinal illuminance as a function of pupil size (Thibos, Lopez-Gil and Bradley, 2018). Although Troland accounts for the pupil area $A_P$ in Equation (10), one still needs to make an assumption about the eye's focal distance $f_e$ (i.e. use a value for a standard eye) to calculate an actual troland value (Bass, Enoch and Lakshminarayanan, 2010, p. 185). In our derivation however, $f_e$ cancels out in Equation (11) after setting $E_{R_{Lam}}$ equal to $E_{R_{Scan}}$. This means that $L_{RS}$ is independent from any assumptions about the eye apart from the pupil size.

We picked the model from Stanley & Davies (Stanley and Davies, 1995) to express pupil size as a function of ambient luminance. Using another model with additional parameters such as age (Watson and Yellott, 2012) could further increase the accuracy but would also increase the complexity. The fundamental relationship between pupil size and RS luminance holds independently from this model choice.

Using the ambient luminance $L_A$ and combiner transmittance $\tau_C$ we defined the ambient contrast ratio for RS glasses as $ACR_{RS} = \frac{L_{RS}}{\tau_C L_A}$. In some previous research (Fidler *et al.*, 2021; Xiong *et al.*, 2021), the ambient contrast ratio has been defined as $ACR = \frac{L_{on} + \tau L_A}{L_{off} + \tau L_A}$ where $L_{on}$ and $L_{off}$ are the luminance of on- and off-pixels, $\tau$ is the see-through transmittance and $L_A$ is the ambient luminance. RS displays do not project light into dark pixel areas. $L_{off}$ is always zero so the denominator of the two formulas does not differ. More thought must be put into the ambient light term $\tau L_A$ in the numerator. If the term is included, the ambient contrast ratio would equal 1 for an off display. This seems in conflict with the conventional notion of a contrast ratio. We therefore propose to make a distinction between a standard "ambient contrast ratio" that does not include the background term $\tau L_A$ in the numerator and a second quantity, potentially called the "additive ambient contrast ratio", that does include it. In this paper we use the standard ambient contrast ratio.

With the derivation for the brightness and contrast of visual stimuli we were finally able to design and evaluate our psychophysical study for the perceived resolution. Overall, we found consistent results across 15 trials with different stimuli parameters. The study had a gamified design, but we provided no reward for correct answers. Instead, we instructed the participants to give accurate feedback according to their perception. Several participants reported after the study that they did not find any visual cues to manipulate the results even if they had wanted to, hinting towards a robust experimental design.

From the field-of-view size and number of digital pixels we can infer that the pixels were not perfectly quadratic. This led to different spatial frequencies for horizontal and vertical staircase levels (**Table 1**). The fact that the correct response ratios in Fig. 6 still follow the trend of a common overall sigmoidal function further speaks to the validity of our measurement approach.

The 0.8 decimal (0.097 logMAR) minimum visual acuity needed to be included in the study corresponds to a visual resolution of 24.0 cyc/deg (Caltrider, Gupta and Tripathy, 2024). Since we found an average perceived resolution threshold of 7.2 cyc/deg across all trials, we conclude that the perceived resolution was limited by the optical resolution of the RS glasses and not the visual perception of any of the participants.

Our results indicate that the background conditions did not have a notable influence on the perceived resolution within the range of conditions we tested. This could however still be the case for other parameter ranges. We see our study as a first step towards such human-centered characterizations of augmented reality glasses and recognize potential for follow-up studies in additional parameter ranges.

We also measured the RS glasses optical display resolution according to the IEC 61947-2 standard (International Electrotechnical Commission (IEC), 2001) and found a maximum value of 3.4 cyc/deg for monochromatic green images. In our psychophysical study we found a perceived resolution threshold of 7.4 cyc/deg for monochromatic green images.

The perceived resolution threshold adds additional interpretability to the IEC 61947-2 optical resolution threshold, which only uses a modulation depth of 30 % to determine the spatial frequency cutoff, independent of visual perceptibility. Knowing the number of pixels per degree that humans can actually distinguish in the glasses can help in taking better guided design decisions for future prototypes, for example for finding the necessary field-of-view and holographic combiner area size to show image content with a certain desired size and resolution.

Furthermore, in accordance with the ideas presented in section 2.1.2, we expect the ratio between the perceived and optical resolution to be higher for retinal scan displays than for conventional Lambertian displays. A promising next step to test this claim would be to conduct a second, comparative study with a Lambertian augmented reality display that only differs in its optical architecture and is otherwise as identical to our retinal scan glasses prototype as possible.

The BML500P prototype used in this study still had some known limitations. In Fig. 4 one can see brightness inhomogeneities, particularly in the top left corner of the green and blue stimuli. Such inhomogeneities typically arise from local variations in diffraction efficiency of the HOE. This is a manufacturing issue that will be resolved in the future.

The second limitation was the small exit pupil of the single eyebox prototype. The RS display is only visible if the narrow light beam fully passes through the pupil. If this alignment is lost, the image is not visible anymore. Therefore, we added a 5-minute fitting procedure for each participant at the beginning of our study. The nose bridge of the glasses could be quickly adjusted to fit to the pupil position of each participant (see Fig. 1, left). In the future, several approaches such as eyebox replication or eyebox steering can be implemented to remediate this issue.

## 5    Conclusion

We found a physical derivation for the perceived brightness and ambient contrast ratio of holographic augmented reality retinal scan glasses. Building on those results we designed and conducted a psychophysical study to determine the perceived resolution of a fully functional retinal scan glasses prototype.

The perceived brightness as quantified by the retinal scan luminance $L_{RS}$ changes with pupil size: It increases for small pupils and decreases for large pupils. As a result, retinal scan displays appear brighter in bright and darker in dark environments in comparison to conventional displays. This is a unique and advantageous property of retinal scan glasses for augmented reality use cases.

The psychophysical study produced consistent perceived resolution thresholds of about 7 cycles per degree across all trials despite varying background and stimulus conditions. This result shows that within the range of parameters we tested, the optical resolution of the retinal scan glasses and not the human visual perception was the limiting factor for the perceived resolution. Overall, the study demonstrated how both the display and the background can be included in a psychophysical experiment to better understand and optimize augmented reality display performance for realistic use cases.

## 6 Conflict of Interest

MR, PN and SP were employed by Bosch Sensortec GmbH, Reutlingen, Germany during the creation of this work. MR and SP filed patents for Bosch Sensortec GmbH on retinal scan glasses technology. TS declares that the research was conducted in the absence of any commercial or financial relationships that could be construed as a potential conflict of interest.

## 7 Funding

Bosch Sensortec supported this study and itself acknowledges financial support for the development of the augmented reality glasses by the Federal Ministry for Economic Affairs and Climate Action on the basis of a decision by the German Bundestag and by the Ministry for Economic Affairs, Labor and Tourism of Baden-Württemberg based on a decision of the State Parliament of Baden-Württemberg as well as co-financed with tax funds on the basis of the budget financed by the European Union - NextGenerationEU.

## 8 Acknowledgments

We thank Andreas Petersen, Hendrik Specht, Matthias Heberle, Julian Heinzelmann, Henning Kästner and the entire Bosch Sensortec optical engineering team for many fruitful discussions and technical support during this study. We acknowledge support from the Open Access Publishing Fund of the University of Tübingen.

# Appendix

**Table 2.** Mathematical symbols, their meanings and value types for the derivation of the Retinal Scan Luminance and Ambient Contrast Ratio. Final values are the goal of the derivation. Measurement values are measured directly. Intermediate values are used to derive or calculate the final values from the measurement values but are not directly measured themselves. Constant values are physical/photometrical constants.

| Symbol | Meaning | Value type |
|---|---|---|
| $A_P$ | Area of human pupil | Intermediate |
| $A_R$ | Area of image on retina | Intermediate |
| $A_{R_{Lam}}$ | Area of image on retina for a Lambertian system | Intermediate |
| $A_{R_{Scan}}$ | Area of image on retina for a Retinal Scan system | Intermediate |
| $A_S$ | Area of a light source (e.g. area of a display panel) | Intermediate |
| $ACR_{RS}$ | Ambient Contrast Ratio of a Retinal Scan system | Final |
| $d_{P_{mm}}$ | Diameter of human pupil in millimeters | Measurement |
| $E_R$ | Retinal illuminance | Intermediate |
| $E_{R_{Lam}}$ | Retinal illuminance from a Lambertian system | Intermediate |
| $E_{R_{Scan}}$ | Retinal illuminance from a Retinal Scan system | Intermediate |
| $f_e$ | Focal distance of the human eye | Intermediate |
| $K_m$ | Luminous efficacy constant | Constant |
| $L$ | Luminance | Intermediate |
| $L_A$ | Ambient luminance | Measurement |
| $L_{Lam}$ | Luminance of a Lambertian system | Intermediate |
| $L_{RS}$ | Luminance of a Retinal Scan system | Final |
| $R$ | Distance between human pupil and a Lambertian display | Intermediate |
| $V(\lambda)$ | Luminous efficiency function | Constant |
| $\alpha$ | First projection angle of a Retinal Scan system | Measurement |
| $\beta$ | Second projection angle of a Retinal Scan system | Measurement |
| $\lambda$ | Wavelength | Intermediate |
| $\bar{\lambda}_{blue}$ | Average wavelength of blue Retinal Scan system primary | Measurement |
| $\bar{\lambda}_{green}$ | Average wavelength of green Retinal Scan system primary | Measurement |
| $\bar{\lambda}_{red}$ | Average wavelength of red Retinal Scan system primary | Measurement |
| $\tau_C$ | Transparency of Retinal Scan system optical combiner | Measurement |
| $\Phi_{E_{Scan}}$ | Radiant flux from Retinal Scan system | Intermediate |
| $\Phi_{blue}$ | Radiant flux from blue Retinal Scan system primary | Measurement |
| $\Phi_{green}$ | Radiant flux from green Retinal Scan system primary | Measurement |
| $\Phi_{red}$ | Radiant flux from red Retinal Scan system primary | Measurement |
| $\Phi_V$ | Luminous flux | Intermediate |
| $\Phi_{V_{Lam}}$ | Luminous flux from a Lambertian system | Intermediate |
| $\Phi_{V_{Scan}}$ | Luminous flux from a Retinal Scan system | Intermediate |
| $\Omega$ | Solid angle | Intermediate |